\documentclass[aps,twocolumn,a4paper,floatfix,superscriptaddress,pra,longbibliography]{revtex4-2}
\usepackage[toc,page]{appendix}
\usepackage{braket}
\usepackage{epsfig,graphicx,times}
\usepackage{amstext}
\usepackage{amsmath}
\usepackage{amssymb}
\usepackage{mathrsfs}
\usepackage{dcolumn}
\usepackage{bm}
\usepackage[tight]{subfigure}
\usepackage[colorlinks,linkcolor=blue,anchorcolor=blue,citecolor=blue,urlcolor=blue]{hyperref}
\usepackage{color}
\usepackage{siunitx}
\usepackage{appendix}
\usepackage{placeins}
\usepackage{textcomp}
\usepackage{bbold}
\usepackage{soul}
\usepackage{ulem}
\usepackage{float}
\usepackage{verbatim}
\usepackage{minitoc}
\usepackage[dvipsnames]{xcolor}

\usepackage{soul}
\usepackage[framemethod=tikz]{mdframed}
\usepackage{lipsum} 
\usepackage{epstopdf}
\usepackage{CJK}

\begin{document}
	
	\title{Molecular optomechanically-induced transparency}
	\author{Bin Yin}
	\affiliation{Key Laboratory of Low-Dimensional Quantum Structures and Quantum Control of Ministry of Education,\\ Department of Physics and Synergetic Innovation Center for Quantum Effects and Applications,\\ Hunan Normal University, Changsha 410081, China}
	
	\author{Jie Wang}
	\affiliation{Key Laboratory of Low-Dimensional Quantum Structures and Quantum Control of Ministry of Education,\\ Department of Physics and Synergetic Innovation Center for Quantum Effects and Applications,\\ Hunan Normal University, Changsha 410081, China}	
	
	\author{Mei-Yu Peng}
	\affiliation{Key Laboratory of Low-Dimensional Quantum Structures and Quantum Control of Ministry of Education,\\ Department of Physics and Synergetic Innovation Center for Quantum Effects and Applications,\\ Hunan Normal University, Changsha 410081, China}
	
	\author{Qian Zhang}
	\affiliation{Key Laboratory of Low-Dimensional Quantum Structures and Quantum Control of Ministry of Education,\\ Department of Physics and Synergetic Innovation Center for Quantum Effects and Applications,\\ Hunan Normal University, Changsha 410081, China}
	
	\author{Deng Wang}
	\affiliation{Key Laboratory of Low-Dimensional Quantum Structures and Quantum Control of Ministry of Education,\\ Department of Physics and Synergetic Innovation Center for Quantum Effects and Applications,\\ Hunan Normal University, Changsha 410081, China}
	
	\author{Tian-Xiang Lu}
	\email{lu.tianxiang@foxmail.com}
	\affiliation{College of Physics and Electronic Information, Gannan Normal University, Ganzhou 341000, Jiangxi, China}
	\affiliation{Zhejiang Province Key Laboratory of Quantum Technology and Device, School of Physics, and\\ State Key Laboratory for Extreme Photonics and Instrumentation, Zhejiang University, Hangzhou 310027, China}
	
	\author{Ke Wei}
	\email{weikeaep@163.com;}
	\affiliation{Institute for Quantum Science and Technology, College of Science, National University of Defense Technology, Changsha 410073, China}
	
	\author{Hui Jing}
	\email{jinghui73@foxmail.com}
	\affiliation{Key Laboratory of Low-Dimensional Quantum Structures and Quantum Control of Ministry of Education,\\ Department of Physics and Synergetic Innovation Center for Quantum Effects and Applications,\\ Hunan Normal University, Changsha 410081, China}	
	
	\date{\today}
	
	\begin{abstract}
		Molecular cavity optomechanics (COM), characterized by remarkably efficient optomechanical coupling enabled by a highly localized light field and ultra-small effective mode volume, holds significant promise for advancing applications in quantum science and technology. Here, we study optomechanically induced transparency and the associated group delay in a hybrid molecular COM system. We find that even with an extremely low optical quality factor, an obvious transparency window can appear, which is otherwise unattainable in a conventional COM system. Furthermore, by varying the ports of the probe light, the optomechanically induced transparency or absorption can be achieved, along with corresponding slowing or advancing of optical signals. These results indicate that our scheme provides a new method for adjusting the storage and retrieval of optical signals in such a molecular COM device.
	\end{abstract}
	\maketitle
	
	\section{Introduction}
	Cavity optomechanics (COM)~\cite{aspelmeyer2014cavity,metcalfe2014applications}, exploring the interaction between light fields and mechanical vibration, provides a versatile platform for both fundamental physics and high-precision sensing~\cite{aspelmeyer2012quantum,kippenberg2008cavity,brennecke2008cavity,pirkkalainen2015cavity,li2021cavity,zhu2023cavity,xu2024single,PhysRevA.79.042339}. Recent advances in microfabrication have enabled the integration of nanomechanical vibrations, such as molecular systems with rich degrees of freedom, into COM devices, facilitating effective coupling with photons~\cite{flick2018strong,galego2015cavity,toninelli2021single}. Molecular COM systems~\cite{schmidt2017linking,roelli2016molecular,esteban2022molecular,benz2016single,BozhevolnyiMortensen+2017+1185+1188,roelli2024nanocavities,li2024magnetic,willets2007localized,patra2023molecular,xu2022phononic}, featuring strong optomechanical coupling between molecular vibrations and light fields with coupling rates ranging from 10 to 100 GHz~\cite{anderson2018two,roelli2016molecular} — approximately five orders of magnitude higher than those in conventional COM, have enabled significant advancements with a wide range of applications, including heat transfer~\cite{ashrafi2019optomechanical}, collective effects and entanglement~\cite{zhang2020optomechanical,huang2024collective}, frequency  up-conversion~\cite{zou2024amplifying,chen2021continuous,xomalis2021detecting}, nonlinear effect~\cite{benz2016single,lombardi2018pulsed}, mechanical lasing~\cite{schmidt2024molecular}, measurement~\cite{liu2018room,liu2017coupled,tabatabaei2015tunable}, and molecule-protein analysis~\cite{sadhanasatish2023molecular}. In recent experiments~\cite{jakob2023giant,boehmke2024uncovering}, giant optomechanical spring effect and molecular low-frequency vibrations were discovered by the surface-enhanced Raman scattering (SERS). Moreover, hybrid molecular COM systems have also emerged by integrating metal nanoparticles deposited on various cavities such as microdisk cavities~\cite{shlesinger2021integrated}, Fabry-P\'erot (F-P) cavities~\cite{shlesinger2023hybrid}, and photonic crystals~\cite{dezfouli2019molecular,abutalebi2024single}, achieving narrower resonances than typical molecular vibrational frequencies and thereby facilitating entry into the sideband-resolved region~\cite{thakkar2017sculpting,pan2019elucidating,kamandar2017modal,gurlek2018manipulation,barreda2021hybrid,lu2022unveiling,li2023highly}. Recent experiments demonstrate that these systems can selectively and independently enhance a single Raman line and facilitate the design of an image sensor-based miniature computational spectrometer~\cite{shlesinger2023hybrid,zhang2024miniature}.
	
	In parallel, the COM system, as a stationary quantum system with a potentially long coherence time~\cite{aspelmeyer2014cavity,metcalfe2014applications,barzanjeh2022optomechanics,kleckner2008creating}, offers an alternative implementation of optical quantum memories or quantum cryptography over long distances~\cite{lvovsky2009optical,lei2023quantum}. Typically, optomechanical memories based on the fast- and slow-light effects are achieved through optomechanically induced transparency (OMIT)~\cite{weis2010optomechanically,safavi2011electromagnetically}, in which the abnormal dispersion can alter the light group velocity. OMIT has been experimentally demonstrated in various optomechanical systems~\cite{burek2016diamond,teufel2011circuit,safavi2011electromagnetically,dong2014optomechanically,shen2016compensation}, providing the basis for a wide range of applications, e.g. optical communications~\cite{zhou2013slowing,tang2022quantum}, high sensitivity sensors~\cite{xiong2017highly,kong2017coulomb,xiong2017precision}, and mechanical cooling~\cite{ojanen2014ground}. Also, many novel OMIT phenomena have been studied theoretically, such as nonreciprocal OMIT~\cite{lu2017optomechanically,sun2024multicolor,zhang2024nonreciprocal}, non-Hermitian OMIT~\cite{lu2018optomechanically,pan2024optomechanically,jing2015optomechanically,peng2023nonreciprocal}, and nonlinear enhanced OMIT~\cite{lu2023magnon,jiao2018optomechanical,li2016transparency,Lu:24}. Recently, OMIT-based memories have demonstrated in silica microspheres~\cite{fiore2011storing}, diamond microdisks~\cite{lake2021processing}, and F-P cavities with a soft-clamped membrane show promising potential for applications in modern optical and future quantum networks~\cite{kristensen2024long}. However, as far as we know, the generation and coherent control of OMIT in a molecular COM system have not been explored.
	
	In this work, we investigated the OMIT of molecular COM formed by the metal nanoparticle and a large number of molecules deposition on a microdisk cavity, mainly including the transmission rate and optical group delay of the probe field. We show that (i) the OMIT spectra of molecular COM exhibit more obvious transparency peaks compared to conventional COM, owing to the enhanced optomechanical coupling strength and the collective effects of molecules; (ii) the transparency and absorption of optical signals can be adjusted by varying the ports of the probe light; and (iii) as a result, tunable slow or fast light can be achieved, which has led to practical applications, such as in the selective storage and retrieval of optical signals. These features show that hybrid molecular COM devices can serve as powerful tools for manipulating photons and phonons, with potential applications in optical signal storage and communications, paving the way for more discussion in molecular COM, even in quantum metrology.
	
	The paper is organized as follows. In Sec.~\ref{Sec:ModelandHamiltonian}, we introduce our model of a molecular COM with the microdisk cavity and calculate the transmission rate and group delay of the probe light. Furthermore, in Sec.~\ref{Sec: Results and discussions}, we provide a detailed discussion of the advantages of this system compared to conventional COM, along with the storage and retrieval effects for optical signals in this device. Finally, we present a summary and outlook in Sec.~\ref{Sec:Conclusions}.
	
	\begin{figure}[t]
		\centering\includegraphics[width=0.95\linewidth]{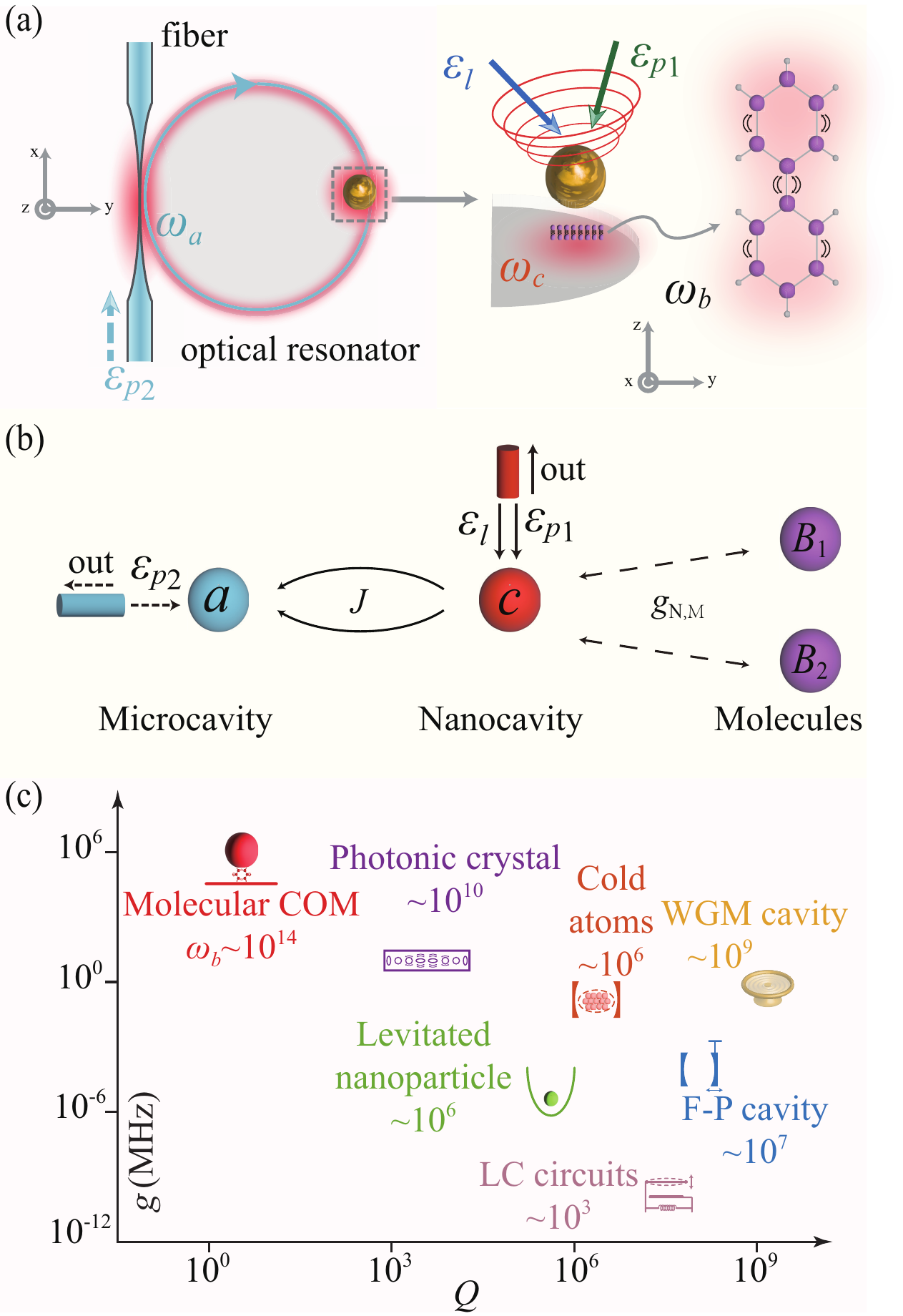}
		\caption{(a) Schematic diagram of the hybrid molecular COM system. The metallic nanoparticle and the molecules (with frequency $\omega_b$) deposit on the microdisk cavity (with frequency $\omega_a$) to form a plasmonic nanocavity (with frequency $\omega_c$ ). Top view of the system model (left), front view of the deposition location of metal nanoparticle (middle), and molecular vibration diagram (right). (b) Schematic of the equivalent mode-coupling model. The plasmonic cavity mode is coupled to $N$ molecular vibrational modes via optomechanical interactions and to the microdisk cavity mode through evanescent fields~\cite{esteban2022molecular,shlesinger2021integrated}. (c) Comparison of optomechanical coupling strength $g$, optical quality factor $Q$, and vibrational frequency $\omega_b$ characterizing different types of optomechanical systems, including molecular COM~\cite{roelli2016molecular,verhagen2012quantum,murch2008observation,akahane2005fine,liu2011high,bagci2014optical,thompson2008strong}.}
		\label{fig1}
	\end{figure}
	
	\section{Theoretical model}\label{Sec:ModelandHamiltonian}
	As shown in Fig.~\ref{fig1}(a), we consider a hybrid molecular COM system that comprises molecules within a plasmonic nanocavity coupled to a microdisk cavity~\cite{thakkar2017sculpting,pan2019elucidating,shlesinger2023hybrid}.  The microdisk cavity couples to an optical waveguide, and supports a whispering-gallery-mode with resonance frequency $\omega_a$, effective cavity mode volume $V_a$, and decay rate $\kappa_a$. Simultaneously, the metallic nanoparticle and the molecules co-deposit on the microdisk cavity to construct the nanoparticle-on-cavity (NPoC) structure, thereby forming a plasmonic nanocavity~\cite{shlesinger2021integrated,shlesinger2023hybrid,dezfouli2019molecular,kristensen2014modes}. The plasmonic nanocavity couples to the microdisk cavity field through the evanescent fields with the coupling strength $J$, and supports an optical mode with resonance frequency $\omega_c$, effective cavity mode volume $V_c$, and decay rate $\kappa_c$. A large number of vibration molecules is placed in the gap between the metal nanoparticle and the microdisk cavity. In the process of depositing molecules in experiments~\cite{chen2021continuous,boehmke2024uncovering,jakob2023giant,benz2016single},  it cannot be guaranteed that molecules will always be an ensemble of molecules. Therefore, considering two or more ensembles of molecules is a general situation, that can always be encountered in experiments. As a demonstration, we only consider the simple case here, which is two molecular ensembles with resonance frequency $\omega_b$, decay rate $\gamma$~\cite{roelli2016molecular,esteban2022molecular,zhang2020optomechanical,huang2024collective}. The coupling between vibrational and plasmonic nanocavity mode is purely parametic, and thus the single photon optomechanics interaction is $-\hbar gc^{\dagger}c(b^{\dagger}+b)$ [see Fig.~\ref{fig1}(b)], in which $g$$\,=\,$$-\omega_c R_b \sqrt{\hbar / 2 m \omega_b} /\left(\varepsilon_0 V_c\right)$, where $b$ $(b^{\dagger})$ is the annihilation (creation) operator for the vibrational mode of single molecule, the general value of $V_c$ is $\sim2.5\times10^{-7}\,\mathrm{\mu m^3}$, and $R_b$ is the Raman polarizability with typical values $\sim3\times10^{-10}\varepsilon^{2}_{0}$ {\AA}$^{4}$amu$^{-1}$. $\omega_b$ is typically between $5\sim50\,\mathrm{THz}$, and the optomechanical coupling coefficient $g$ is estimated on the order between $2\pi\times10\sim2\pi\times100\,\mathrm{GHz}$, as shown in Fig.~\ref{fig1}(c)~\cite{roelli2016molecular,esteban2022molecular}. To explore the response of the optical signals in different channels, a strong pump light is applied at frequency $\omega_{l}$ to drive the system from plasmonic nanocavity and a weak probe light at frequency $\omega_{p}$ to drive the system from the microdisk cavity or plasmonic nanocavity. In a frame rotating $U$$\,=\,$$e^{-i\omega_{l}t(a^{\dagger}a+c^{\dagger}c)}$, the Hamiltonian of this system can be written as
	\begin{eqnarray}
		\begin{aligned}
			H =&H_{\mathrm{0}}+H_{\mathrm{I}}+H_{\mathrm{dr}}, \\ H_{0} =&\hbar\Delta_aa^{\dagger}a+\hbar\Delta_{c}c^{\dagger}c+\sum_{j}^{N}\hbar\omega_{b}b_{j}^{\dagger}b_{j}+\sum_{k}^{M}\hbar\omega_{b}b_{k}^{\dagger}b_{k}, \\ H_{\mathrm{I}} =&-\sum_{j}^{N}\hbar gc^{\dagger}c(b_{j}^{\dagger}+b_{j})-\sum_{k}^{M}\hbar gc^{\dagger}c(b_{k}^{\dagger}+b_{k})\\&+\hbar J(a^{\dagger}c+ac^{\dagger}), \\ H_{\mathrm{dr}} =&i\hbar(\epsilon_{l}c^{\dagger}+\epsilon_{p1}c^{\dagger}e^{-i\Delta_{p}t}+\epsilon_{p2}a^{\dagger}e^{-i\Delta_{p}t})-\mathrm{H.c.},\end{aligned}
		\label{1}
	\end{eqnarray} 
	Our motivation is to investigate the impact of uncontrollable factors in this experiment, studying the effects of molecules from different ensembles ($B_1$ and $B_2$) on OMIT, as well as the corresponding group delay. To this end, one defines the two mechanical modes through the collective effect of molecules $B_{1}=\sum_{i=1}^{N}b_{j}/\sqrt{N}$($B_{2}=\sum_{i=1}^{M}b_{k}/\sqrt{M}$)~\cite{zhang2020optomechanical,huang2024collective,sun2003quasi,emary2003quantum}. Thus, the Hamiltonian $H$ can be rewritten as
	\begin{eqnarray}
		\begin{aligned}
			H =&H_{\mathrm{0}}+H_{\mathrm{I}}+H_{\mathrm{dr}}, \\ H_{0} =&\hbar\Delta_{a}a^{\dagger}a+\hbar\Delta_{c}c^{\dagger}c+\hbar\omega_{b}B_{1}^{\dagger}B_{1}+\hbar\omega_{b}B_{2}^{\dagger}B_{2}, \\ H_{\mathrm{I}} =&-\hbar g_{N}c^{\dagger}c(B_{1}^{\dagger}+B_{1})-\hbar g_{M}c^{\dagger}c(B_{2}^{\dagger}+B_{2})\\&+\hbar J(a^{\dagger}c+ac^{\dagger}), \\ H_{\mathrm{dr}} =&i\hbar(\epsilon_{l}c^{\dagger}+\epsilon_{p1}c^{\dagger}e^{-i\Delta_{p}t}+\epsilon_{p2}a^{\dagger}e^{-i\Delta_{p}t})-\mathrm{H.c.},\end{aligned}
		\label{1}
	\end{eqnarray} 
	where $\Delta_{a,c}$$\,=\,$$\omega_{a,c}-\omega_{l}$, $\Delta_p$$\,=\,$$\omega_{p}-\omega_{l}$, and $g_N$$\,=\,$$g\sqrt{N}$ ($g_M$$\,=\,$$g\sqrt{M}$) denotes the optomechanical coupling coefficient caused by collective effect. $a$, $c$, and $B_{1,2}$ are the annihilation bosonic field operators describing the microdisk cavity, plasmonic nanocavity, and mechanical modes, respectively. 	$\epsilon_{l}$$\,=\,$$\sqrt{{2P_{l}\kappa_{ex1}}/{\hbar\omega_{l}}}$ ($\epsilon_{p1,2}$$\,=\,$$\sqrt{{2P_{p}\kappa_{ex1,2}}/{\hbar\omega_{p}}}$) is the amplitude of the strong pump (weak probe) field, with the input power of the pump (probe) field $P_{l}$ ($P_{p}$) and the external decay rate $\kappa_{ex1,2}$~\cite{weis2010optomechanically,safavi2011electromagnetically,zhang2018loss}.
	
	Compared to conventional COM systems~\cite{aspelmeyer2014cavity,metcalfe2014applications}, this molecular COM system possesses several notable advantages: (i) 	the significantly higher optomechanical coupling strength, attributed to the strong localization ability of plasmonic nanocavity to light field and its extremely small effective mode volume; and (ii) the substantially higher vibration frequency and collective effect due to the unique characteristics of the molecule. Analogous to previous OMIT works~\cite{safavi2011electromagnetically,weis2010optomechanically}, we focus on the mean response of the system to the probe field without including quantum fluctuation. By classically introducing loss terms to the equations, we obtain the Heisenberg-Langevin equation of the system:
	\begin{eqnarray}
		\begin{aligned}
			\dot{a} =&-(i\Delta_{a}+\kappa_{a})a-iJc+\epsilon_{p2}e^{-i\Delta_{p}t}, \\ \dot{c} =&-(i\Delta_{c}+\kappa_{c})c+ig_{N}c(B_{1}^{\dagger}+B_{1})\\ &+ig_{M}c(B_{2}^{\dagger}+B_{2})-iJa+\epsilon_{l}+\epsilon_{p1}e^{-i\Delta_{p}t}, \\ \dot{B_{1,2}} =&-(i\omega_{b}+\gamma)B_{1,2}+ig_{N,M}c^{\dagger}c.\end{aligned}
		\label{2}
	\end{eqnarray} 
	
	For the OMIT process, the probe light is generally much weaker than the pump light~\cite{lu2023magnon}. Therefore, we can extend each operator to its steady-state mean and first-order perturbation, that is $A$$\,=\,$$A_{s}+\delta A$ $(A$$\,=\,$$a,B_1,B_2,c)$, and we can take the probe light as a perturbation~\cite{safavi2011electromagnetically,weis2010optomechanically}. In this way, the steady-state solutions of the system are easily obtained as
	\begin{eqnarray}
		\begin{aligned}
			a_{s} &=\frac{-iJc_{s}}{i\Delta_{a}+\kappa_{a}}, \\ B_{1s,2s} &=\frac{ig_{N,M}|c_u{s}|^{2}}{i\omega_{b}+\gamma}, \\ c_{s} &=\frac{-iJa_{s}+\epsilon_{l}}{i\Delta_s+\kappa_{c}},\end{aligned}
		\label{3}
	\end{eqnarray}
	where $\Delta_{s}$$\,=\,$$\Delta_{c}-g_{N}(B_{1,s}^{*}+B_{1,s})-g_{M}(B_{2,s}^{*}+B_{2,s})$ displaying the optical frequency shift caused by molecular vibration. Now, when the input port of the probe field and output port with the plasmonic nanocavity, i.e., $\epsilon_{p2}$$\,=\,$$0$ [see the blue dashed line arrow in Fig.~\ref{fig1} (a)], by substituting the expression of the operator into the Eq.~(\ref{2}) and combining steady-state solutions, the perturbation motion equation is obtained as
	\begin{eqnarray}
		\begin{aligned}
			\dot{\delta a} =&-(i\Delta_{a}+\kappa_{a})\delta a-iJ\delta c, \\ \dot{\delta B_{1,2}} =&-(i\omega_b+\gamma)\delta B_{1,2}+ig_{N,M}(c_{s}^{*}\delta c+c_{s}\delta c^{*}), \\ \dot{\delta c} =&-(i\Delta_s+\kappa_{c})\delta c+ig_{N}c_{s}(\delta B_{1}^{*}+\delta B_{1})\\ &+ig_{M}c_{s}(\delta B_{2}^{*}+\delta B_{2})-iJ\delta a+\epsilon_{p1}e^{-i\Delta_{p}t}.\\\end{aligned}
		\label{4}
	\end{eqnarray} 
	
	To explore the characteristics of the OMIT process in this system, we use the following ansatz to solve the amplitudes of the first-order sidebands, which is $\delta A=A^{-}e^{-i\Delta_p t}+A^{+}e^{i\Delta_p t}$~\cite{weis2010optomechanically}. Then substituting ansatz into Eq.~(\ref{4}) leads to one group equations describing the linear response
	\begin{eqnarray}
		\begin{aligned}
			h_{1}^{+}a^{-} =&-iJc^{-},~~~ h_{1}^{-}a^{+*} =iJc^{+*}, \\ h_{2}^{+}c^{-} =&-iJa^{-}+ig_{N}c_{s}(B_1^{+*}+B_1^{-})\\ &+ig_{M}c_{s}(B_2^{+*}+B_2^{-})+\epsilon_{p1}, \\ h_{2}^{-}c^{+*} =&iJa^{+*}-ig_{N}c_{s}^{*}(B_1^{+*}+B_1^{-})\\ &-ig_{M}c_{s}^{*}(B_2^{+*}+B_2^{-}), \\ h_{3}^{+}B_{1,2}^{-} =&ig_{N,M}(c_{s}^{*}c^{-}+c_{s}c^{+*}), \\ h_{3}^{-}B_{1,2}^{+*} =&-ig_{N,M}(c_{s}^{*}c^{-}+c_{s}c^{+*}).\end{aligned}
		\label{5}
	\end{eqnarray} 
	
	Solving Eq.~(\ref{5}) leads to
	\begin{eqnarray}
		\begin{aligned}
			c^{-} &=\frac{h_{1}^{+}(\Pi\Gamma^{-}+2ih_{1}^{-}\lambda)\epsilon_{p1}}{\Pi\Gamma^{+}\Gamma^{-}+2i\lambda(h_{1}^{-}\Gamma^{+}-h_{1}^{+}\Gamma^{-})]},\end{aligned}
		\label{6}
	\end{eqnarray} 
	where 
	\begin{eqnarray}
		\begin{aligned}
			h_{1}^{\pm} &=\pm i\Delta_{a}+\kappa_{a}-i\Delta_{p},h_{2}^{\pm}=\pm i\Delta_s+\kappa_{c}-i\Delta_{p}, \\ h_{3}^{\pm} &=\pm i\omega_{b}+\gamma-i\Delta_{p}, \\ \Gamma^{\pm} &=h_{1}^{\pm}h_{2}^{\pm}+J^{2},\Pi=h_{3}^{+2}h_{3}^{-2}, \\ \lambda &=h_{3}^{+}h_{3}^{-}\omega_b(g_{N}^{2}+g_{M}^{2})|c_{s}|^{2}.\end{aligned}
		\label{7}
	\end{eqnarray} 
	
	By using the standard input-output relations~\cite{gardiner1985input}, i.e., $c_{\mathrm{out}}$$\,=\,$$c_{\mathrm{in}}-\sqrt{\kappa_{ex1}}c(t)$, where $c_{\mathrm{in}}$ ($c_{\mathrm{out}}$) is the input (output) probe operators, the transmission rate of the probe field applied in the plasmonic nanocavity can be written as
	\begin{eqnarray}
		\begin{aligned}
			T_c=|t_{p}|^{2}=|1-\frac{\kappa_{ex1}c^{-}}{\epsilon_{p1}}|^{2}.\end{aligned}
		\label{8}
	\end{eqnarray} 
	in which we set $\kappa_{ex1}=\kappa_c$ for the critically coupled case~\cite{weis2010optomechanically,safavi2011electromagnetically}.
	\begin{figure*}[t]
		\centering
		\includegraphics[width=1\textwidth]{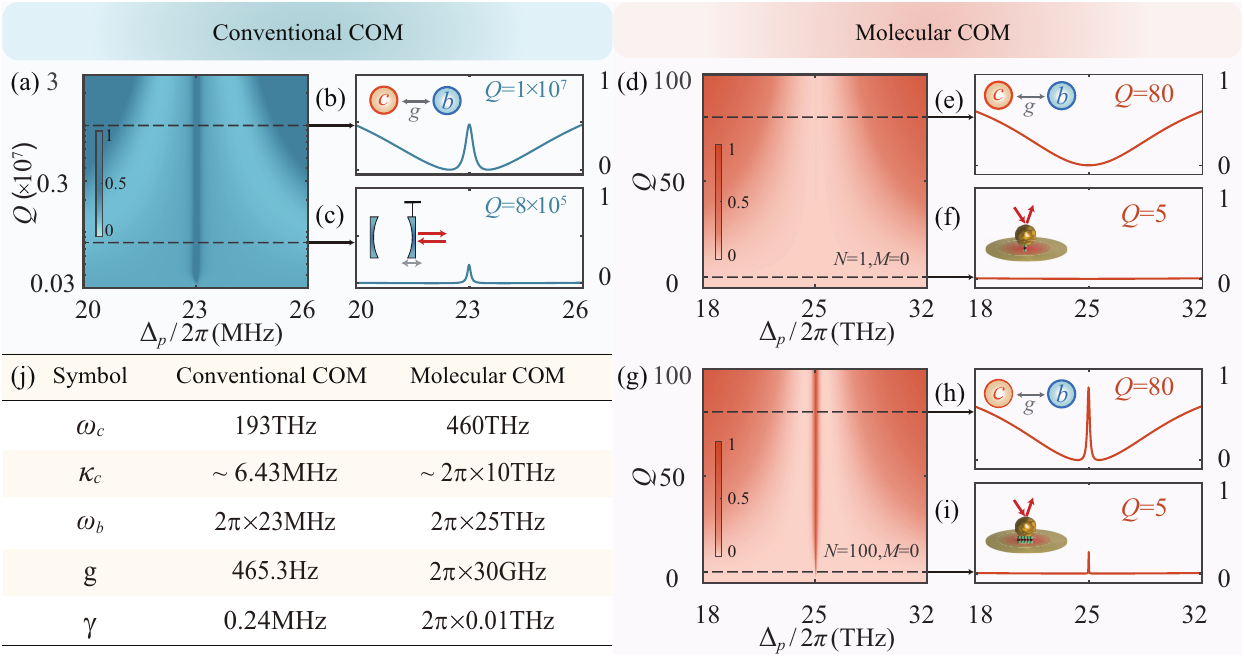}
		\caption{Transmission rate of the probe field as a function of the optical detuning $\Delta_p$ and the optical quality factor $Q$ in the single-cavity case of conventional COM (a) and the molecular COM with one (d) or a hundred molecules (g). As indicated by the markings, transmission rate versus $\Delta_p$ at $Q$$\,=\,$$10^7$ and $Q$$\,=\,$$8\,$$\times$$\,10^5$ for conventional COM (b-c)~\cite{weis2010optomechanically}, and at $Q$$\,=\,$$80$ and $Q$$\,=\,$$5$ for molecular COM with one molecule(e-f)~\cite{he2023single} and with a hundred molecules (h-i). Inset: the corresponding model diagrams. The experimental feasibility parameters are shown in (j) and $P_l$$\,=\,$$1\,\mathrm{mW}$ and $J$$\,=\,$$0$~\cite{roelli2016molecular,shlesinger2021integrated,shlesinger2023hybrid,zhang2018loss}.}
		\label{fig2}
	\end{figure*}
	Similarly, for the input port of the probe field and output port with the microdisk cavity, $\epsilon_{p1}$$\,=\,$$0$ [see the green solid line arrow in Fig.~\ref{fig1} (a)], we can derive the transmission rate of the probe field using the method consistent with the previous situation. Differently, we have input-output relations, i.e., $a_{\mathrm{out}}$$\,=\,$$a_{\mathrm{in}}-\sqrt{\kappa_{ex2}}a(t)$, where $a_{\mathrm{in}}$ ($a_{\mathrm{out}}$) is the input (output) probe operators and $\kappa_{ex2}=\kappa_a$ for critical coupling in this case (see Appendix A for detailed derivation)
	\begin{eqnarray}
		\begin{aligned}
			a^{-} &=\frac{[(h_{2}^{+}\Pi-2i\lambda)(\Pi\Gamma^{-}+2ih_{1}^{-}\lambda)-4h_{1}^{-}\lambda^{2}]\epsilon_{p2}}{(\Pi\Gamma^{+}-2ih_{1}^{+}\lambda)(\Pi\Gamma^{-}+2ih_{1}^{-}\lambda)-4h_{1}^{+}h_{1}^{-}\lambda^{2}}, \\ T_a &=|t_{p}|^{2}=|1-\frac{\kappa_{ex2}a^{-}}{\epsilon_{p2}}|^{2},\end{aligned}
		\label{9}
	\end{eqnarray} 
	
	In this section, we mainly discussed the model structure of this system and derived the transmission rate under different paths. This has made a great contribution to our upcoming discussion.
	
	\begin{figure}[t]
		\centering
		\includegraphics[width=0.95\linewidth]{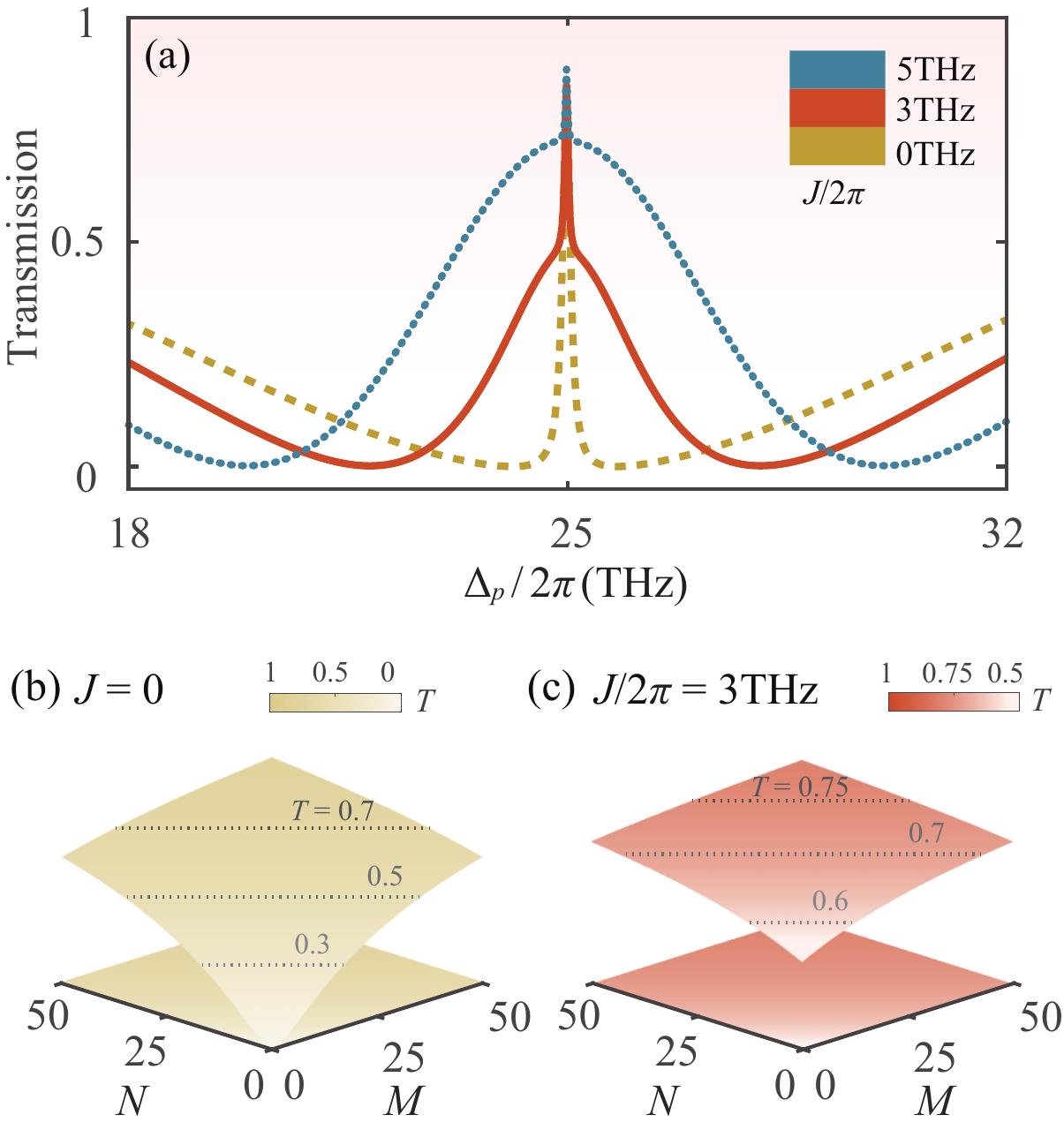}
		\caption{(a) Transmission rate of the probe field as a function of the optical detuning $\Delta_p$ at the different coupling strength $J$ in the molecular COM. Transmission rate as a function of the number of molecules $N$ and $M$ at $\Delta_p/\omega_b$$\,=\,$$1$ for the single-cavity (b) and double-cavities (c) cases of molecular COM. The parameters used here are $N$$\,=\,$$100$ and $M$$\,=\,$$0$ at (a) and $P_l$$\,=\,$$1\,\mathrm{mW}$ (a-c).}
		\label{fig3}
	\end{figure}
	\section{Results and discussions}\label{Sec: Results and discussions}
	\subsection{Single-cavity case}
	In this section, we study the OMIT profile in single-cavity cases for both the conventional COM system and the molecular COM system. In Fig.~\ref{fig2}, the transmission rate of two systems is shown as a function of optical detuning $\Delta_p$$\,=\,$$\omega_p-\omega_l$ and optical quality factor $Q$. For comparison, we first consider the conventional COM system~\cite{weis2010optomechanically}. Under the resonance condition $\Delta_p$$\,=\,$$\omega_b$, the standard OMIT spectrum appears [see Fig.~\ref{fig2} (a)], with the transmission rate decreasing accordingly as the quality factor Q decreases. This phenomenon is caused by interference effects between two absorption channels~\cite{weis2010optomechanically,PhysRevA.86.013815}. It is worth noting that the transparency window is very obvious when $Q$$\,=\,$$10^7$, while the transparency window is closed when $Q$$\,=\,$$8\,$$\times$$\,10^5$, as shown in Figs.~\ref{fig2} (b) and \ref{fig2} (c). This is due to the characteristics of the conventional COM, where the single photon optomechanical coupling strength is relatively weak. Therefore, to achieve a window, the support of a high-quality factor $Q$ is necessary.
	
	For the molecular COM system with only one molecule, it is difficult to achieve the transparent window [see Figs.~\ref{fig2} (d-f)] due to the extremely low optical quality factor $Q$$\,=\,$$1$$\,\sim\,$$100$. However, in the case of multiple molecules~\cite{zhang2020optomechanical,huang2024collective}, the collective effects can significantly enhance the effective optomechanical coupling, thus achieving an obvious transparency window [see Figs.~\ref{fig2} (g-i)]. Notably, transparent windows can still appear even at extremely low optical quality factors, i.e., $Q$$\,=\,$$5$ [see Fig.~\ref{fig2} (i)]. This is precisely a manifestation of the enormous advantages of the molecular COM. Overall, the OMIT spectra of the two systems show significant similarities. However, conventional COM necessitates an exceptionally high-quality factor ($Q\sim10^7$), whereas molecular COM can achieve OMIT spectra at much lower $Q$$\,=\,$$1$$\,\sim\,$$100$ values by leveraging their strong optomechanical coupling and collective effects. 
	
	\subsection{Double cavities case}
	Hybrid molecular COM system, formed by coupling a plasmonic nanocavity to the microdisk cavity, has the advantages of high $Q$ and low mode volume~\cite{shlesinger2021integrated,shlesinger2023hybrid,dezfouli2019molecular,abutalebi2024single}. Here, compared to the single-cavity case, we mainly focus on the influence of the assistance of microdisk cavity on the transmission spectrum. The transmission rate is shown in Fig.~\ref{fig3} (a) as a function of optical detuning $\Delta_p$ under different optical couplings $J$. In comparison to the standard OMIT spectrum (see the yellow dashed line) in a single cavity, a cascaded transparency window can be observed in the hybrid molecular COM system. Upon sweeping the probe field across the plasmonic nanocavity, a transparency window within the broad cavity photon resonance is initially observed due to photon-photon coupling under the condition where $\kappa_c$$\,\gg\,$$ J$. Concurrently, a second transparency window with a linewidth equal to the decay rate of the mechanical mode can be witnessed as a consequence of the optomechanical coupling. This indicates that, with the augmentation of the coupling strength $J$, the exchange of photons between the two cavities is expedited, thereby further attaining a higher transmission rate.
	
	In addition, we further discussed more general molecular collective effects in both single-cavity and double-cavities cases. For $J$$\,=\,$$0$ [see Fig.~\ref{fig3} (b)], when $N$$\,=\,$$0$ and $M$$\,=\,$$0$, it represents a pure optical mode devoid of any interference effect, corresponding to resonant absorption with $T$$\,=\,$$0$~\cite{safavi2011electromagnetically,weis2010optomechanically}. When $N$$\,\neq\,$$0$ and $M$$\,\neq\,$$0$, it can be observed that with the involvement of optomechanical coupling, interference takes place and a transparent window emerges. Moreover, whether with the increase of $N$ or $M$, we can find the transmission rate is linearly related to $\sqrt{N}$ or $\sqrt{M}$. For $J/2\pi$$\,=\,$$3\,\mathrm{THz}$ [see Fig.~\ref{fig3} (c)], when $N$$\,=\,$$0$ and $M$$\,=\,$$0$, interference effects will occur with the assistance of a microdisk cavity, resulting in cascade transparency phenomena. Similarly, whether with the increase of $N$ or $M$, the same linear relationship can be observed, which stems from the collective effect of the molecule $g_N$$\,=\,$$g\sqrt{N}$ and $g_M$$\,=\,$$g\sqrt{M}$~\cite{zhang2020optomechanical,huang2024collective}. As the number of growing molecules increases, we demonstrate OMIT at relatively weak driving power, thereby further avoiding a series of effects caused by extremely high driving power and preventing device damage, which is challenging to achieve in conventional COM. Due to the collective effect of molecules, the effective optomechanical coupling can be enhanced, enabling the system to enter strong coupling or even ultrastrong coupling regions, thus facilitating the study of more novel quantum effects~\cite{forn2019ultrastrong,frisk2019ultrastrong,hu2015quantum,dovzhenko2018light,machado2016quantum,QIN20241}.
	\begin{figure*}[t]
		\centering
		\includegraphics[width=1\textwidth]{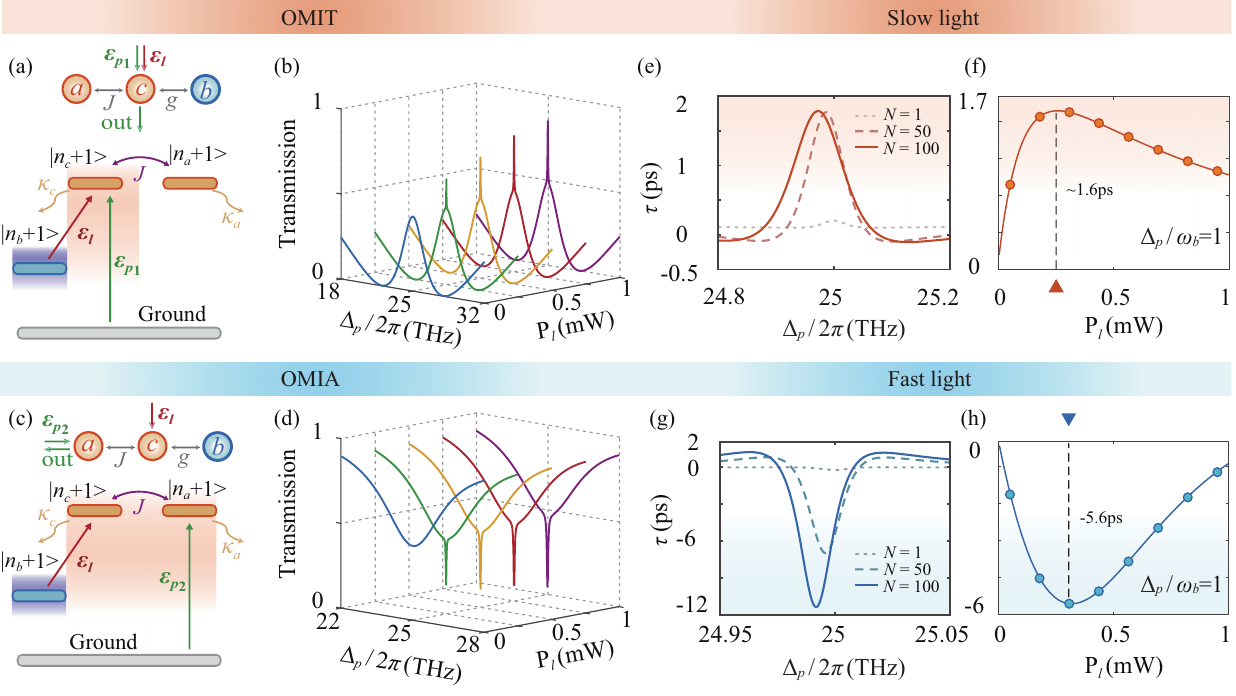}
		\caption{OMIT (a-b) and OMIA (c-d). Transmission rate as a function of the optical detuning $\Delta_p$ and input power of the pump field $P_l$ for the input port of the probe field and output port with the plasmonic nanocavity (b), for the input port of the probe field and output port with the microdisk cavity (d). Corresponding schematic of equivalent mode-coupling model and energy level diagrams of the system in these cases (a) (c). Level $|n_a+1>$, $|n_b+1>$, and $|n_c+1>$ are the excited state of the microdisk cavity, vibration mode, and plasmonic nanocavity, respectively. Slow-light (e-f) and fast-light (g-h). Group delay $\tau$ of the probe field as a function of the optical detuning $\Delta_p$ at different numbers of molecules $N$ and of the input power $P_l$ for the input port of the probe field and output port with the plasmonic nanocavity (e)-(f), for the input port of the probe field and output port with the microdisk cavity (c)-(d). The parameters used here are $N$$\,=\,$$100$ at (a-d), $P_l$$\,=\,$$0.5\,\mathrm{mW}$ at (e)(g) and $J/2\pi$$\,=\,$$3\,\mathrm{THz}$ and $M$$\,=\,$$0$ at (a-h).}
		\label{fig4}
	\end{figure*}
	\vspace{-0.7em}
	\subsection{Tuned transparency and absorption}
	Compared to the single-cavity case of molecular COM, hybrid molecular COM presents an additional waveguide channel. Consequently, we further investigate the characteristics of transmission rate by varying the ports of the probe light, that is, the probe light applied in the plasmonic nanocavity or microdisk cavity which corresponds to $\kappa_{ex1}$$\,=\,$$\kappa_c$ or $\kappa_{ex2}$$\,=\,$$\kappa_a$ in the input-output relationship. We initially consider the scenario where the probe field is sent into the plasmonic cavity mode [see Fig.~\ref{fig4} (a)-(b)]. As shown in Fig.~\ref{fig4} (b), a cascaded transparency window emerges, which is caused by the interference of three absorption paths for probe photons: the first path is between $|n_b,n_c+1>$ and $|n_b+1,n_c>$ induced by the driving field; the second path connects $|n_a,n_c+1>$ and $|n_a+1,n_c>$, resulting from the coupling between these two cavities; the third path arises from the probe field itself [see energy-level structure in Fig.~\ref{fig4} (a)]. Moreover, as the driving light power increases, the effective optomechanical coupling is enhanced~\cite{PhysRevA.86.013815}, leading to an increase in the transmission rate.
	
	In contrast, when the probe light is input from the waveguide and output from the waveguide [see Fig.~\ref{fig4} (c)-(d)], the transmission spectrum exhibits an absorption dip for $P_l$$\,=\,$$0\,\mathrm{mW}$ (as shown by the solid blue line). Even if a three-level system is maintained similar to the previous situation, due to the high optical loss $\kappa_c $$\,\gg\,$$J$$\,\gg\,$$\kappa_a$, the state $|n_a,n_b,n_c+1>$ can decay to $|n_a,n_b,n_c>$ and the exciton in microdisk cavity will tunnel to the plasmonic nanocavity ($|n_a+1,n_c>$$\,\longrightarrow\,$$|n_a,n_c+1>$), resulting in resonant absorption of the probe field. When $P_l$$\,\neq\,$$0$, the path ($|n_b,n_c+1> $$\,\longleftrightarrow\,$$ |n_b+1,n_c>$) is opened. Subsequently, due to the addition of optomechanical coupling, new interference occurs in the system, giving rise to the second absorption dip, that is, an optomechanical-induced absorption (OMIA) spectrum~\cite{zhang2017optomechanically,hou2015optomechanically,zhang2018double}. Moreover, as the driving light power increases, the absorption valley deepens. Consequently, the transition between transmission and absorption at the resonance point can be achieved by altering the input and output ports. This property could prove practically useful for regulating the transmission and absorption of external optical signals ~\cite{xu2015controllable,li2017optical,si2017optomechanically}.
	
	\subsection{Tunable slow or fast light}
	Accompanying the OMIT process, the fast- and slow-light effect can also emerge due to the abnormal dispersion, which is characterized by the group delay~\cite{safavi2011electromagnetically,weis2010optomechanically,jing2015optomechanically,zhou2013slowing},
	\begin{eqnarray}
		\tau=\frac{d\arg(t_{p})}{d \Delta_{p}}.
		\label{13}
	\end{eqnarray}
	
	Based on the discussion in the preceding section, alterations in ports significantly affect the spectrum. This allows for a detailed investigation into the optical group delay characteristics. The group delay $\tau$ of the probe light, as a function of detuning $\Delta_p$ or driving power $P_l$, is depicted in Fig.~\ref{fig4}. For the case where the probe light is sent to the plasmonic cavity mode, slow-light emerges at the resonance point $\Delta_p$$\,=\,$$0$ due to the presence of a transparent window. At this point, the phase and dispersion of the system will change sharply [see Fig.~\ref{fig4} (e)]. Moreover, as the number of molecules increases, the slow-light is enhanced. The group delay varies with the driving light power, maintaining the slow light effect and reaching its maximum value $\tau_{\mathrm{max}}$$\,\sim\,$$1.6\,\mathrm{ps}$ at  $P_l$$\,=\,$$0.25\,\mathrm{mW}$ [see Fig.~\ref{fig4} (f)]. At this time, the system demonstrates the strongest storage effect on external optical signals.
	
	In contrast, when the probe light is input from the waveguide and output from the waveguide, a pronounced fast-light effect emerges at $\Delta_p$$\,=\,$$0$ due to the presence of an absorption valley [see Fig.~\ref{fig4} (g)]. Moreover, as the number of molecules increases, the fast-light effect is intensified. At the resonance point, as the driving light power $P_l$ varies, the group delay maintains the fast-light effect and reaches its minimum value $\tau_{\mathrm{min}}$$\,=\,$$-5.6\,\mathrm{ps}$ at $P_l$$\,=\,$$0.31\,\mathrm{mW}$ [see Fig.~\ref{fig4} (h)]. Consequently, this indicates a novel method of slowing or advancing optical signals by adjusting different paths, providing a mechanism for selective storage and retrieval of external optical signals.
	
	\subsection{Experimental feasibility}\label{Sec:Experimental feasibility}
	Recent experiments have employed molecular COM devices, which are fabricated by depositing metal nanoparticles and various types of molecules—such as biphenyl-4-thiol~\cite{chen2021continuous,boehmke2024uncovering,jakob2023giant}, G-band of graphene~\cite{narula2010absolute}, and carbon nanotubes~\cite{narula2010absolute}—onto a metal mirror. These devices have been utilized to achieve frequency up conversion~\cite{chen2021continuous}, detect the low-frequency vibrations of molecules~\cite{boehmke2024uncovering}, and demonstrate the giant optomechanical spring effect~\cite{jakob2023giant}. In these experiments, the effective mode volume $V_c$ is typically between $2.5\times10^{-7}$$\,\sim\,$$7\times10^{-5}\,\mathrm{\mu m^3}$ which can be adjusted by employing metal nanoparticles of different sizes or shapes and altering the gap between nanoparticles and the gold mirror or cavity; Mechanical mode frequency $\omega_b$ is typically between $5$$\,\sim\,$$50\,\mathrm{THz}$ and also can be adjusted by using different types of molecules; The Raman tensortypically ranges from 5 to $10^{4}$ {\AA}$^{4}$amu$^{-1}$, and the optomechanical coupling coefficient of $g/2\pi$ in the range of 10 to 100 GHz can be achieved.
	
	Notably, a hybrid molecular COM system is constructed by depositing metal nanoparticles and molecules on microdisk cavities, F-P cavities, or photonic crystals~\cite{thakkar2017sculpting,pan2019elucidating,shlesinger2023hybrid,zhang2024miniature}. Based on these hybrid molecular COM systems, diverse phenomena have been investigated, such as discussing the energy exchange of the system by depositing on the WGM cavity~\cite{thakkar2017sculpting,pan2019elucidating}, enhancing selectively and independently a single Raman line~\cite{shlesinger2023hybrid} and facilitating the design of an image sensor-based miniature computational spectrometer~\cite{zhang2024miniature} by depositing on the F-P cavity. In these experiments, the optical quality factor range of the microdisk cavity or Fabry-P\'erot cavity is $Q$$\,=\,$$10^2$$\,\sim\,$$10^6$, and optical coupling strength between the plasmonic cavity and microdisk cavity is within the range of $J/2\pi$$\,=\,$$0.1\sim10\,\mathrm{THz}$. In general, a microdisk cavity resonator can support two degenerate counterpropagating optical modes, the clockwise (CW) and counterclockwise (CCW) modes, which may couple due to defects within the cavity and surface roughness. Moreover, the deposition of metal nanoparticles and molecules on a microdisk may also induce coupling between the two modes. However, recent experiments have confirmed that backscattering-induced coupling can be counteracted by depositing metal nanoparticles in special positions to generate interference effects between the backscatterings~\cite{svela2020coherent}. Furthermore, in experiments, the influence of backscattering can be completely avoided by depositing metal nanoparticles on the F-P cavity ~\cite{shlesinger2023hybrid,li2023highly,zhang2024miniature}.

	\section{Conclusions}\label{Sec:Conclusions}
	In conclusion, we have studied OMIT and optical group delay in a hybrid molecular COM system composed of the metal nanoparticle and a large number of molecules coupling with a microdisk cavity. We find that compared to the conventional COM system, a standard OMIT spectrum in the molecular COM system with an exceptionally low optical quality factor $Q=1-100$ can be achieved. In addition, we find that by changing the ports of the probe light, we can achieve the OMIT or OMIA profile, resulting in the corresponding slowing or advancing of optical signals. These results provide effective and flexible methods for tuning the storage and retrieval of optical signals through this system. In future work, we aim to explore new phenomena such as entanglement~\cite{vitali2007optomechanical,jiao2020nonreciprocal,liu2023phase,ghobadi2014optomechanical,karg2020light,PhysRevResearch.4.033022,PhysRevApplied.22.064001}, photon blockade~\cite{birnbaum2005photon,faraon2008coherent,rabl2011photon,nunnenkamp2011single,huang2018nonreciprocal,liao2013photon,zuo2024chiral,bin2018two,su2022nonlinear,PhysRevApplied.17.054004}, phonon lasing~\cite{grudinin2010phonon,zhang2018phonon,jing2014pt,jiang2018nonreciprocal,lu2024quantum,zhang2018phase}, and ultrasensitive quantum sensing~\cite{gavartin2012hybrid,wang2024quantum,zhao2020weak,zhang2024quantum,Bin:19}, enabled by the high optomechanical coupling strength and molecular collective effects of the molecular COM system.

	\begin{acknowledgements}
		H.J. is supported by the National Natural Science Foundation of China (NSFC) (Grant No. 11935006, 12421005), the National Key R\&D Program (2024YFE0102400), the Hunan Major Sci-Tech Program (2023ZJ1010). T.-X.L. is supported by the NSFC (12205054).
	\end{acknowledgements}

	
	\appendix
	\renewcommand{\thesubsection}{\thesection\arabic{subsection}}
	\section{Derivation of transmission rate}
	Now, considering the perturbation caused by the probe field, the perturbation equation of motion is obtained by eliminating steady-state values for the input port of the probe field and output port with the microdisk cavity, $\epsilon_{p1}$$\,=\,$$0$ [see the green solid line arrow in Fig.~\ref{fig1} (a)],
	\begin{eqnarray}
		\begin{aligned}
			\dot{\delta a} =&-(i\Delta_{a}+\kappa_{a})\delta a-iJ\delta c+\epsilon_{p2}e^{-i\Delta_{p}t}, \\ \dot{\delta c} =&-(i\Delta_s+\kappa_{c})\delta c+ig_{N}c_{s}(\delta B_{1}^{*}+\delta B_{1})\\ &+ig_{M}c_{s}(\delta B_{2}^{*}+\delta B_{2})-iJ\delta a, \\ \dot{\delta B_{1,2}} =&-(i\omega_b+\gamma)\delta B_{1,2}+ig_{N,M}(c_{s}^{*}\delta c+c_{s}\delta c^{*}).\end{aligned}
		\label{A1}
	\end{eqnarray} 
	
	Substituting ansatz into Eq.~(\ref{A1}) leads to one group equation describing the linear response
	\begin{eqnarray}
		\begin{aligned}
			h_{1}^{+}a^{-} =&-iJc^{-}+\epsilon_{p2},~~~ h_{1}^{-}a^{+*} =iJc^{+*}, \\ h_{2}^{+}c^{-} =&-iJa^{-}+ig_{N}c_{s}(B_1^{+*}+B_1^{-})\\ &+ig_{M}c_{s}(B_2^{+*}+B_2^{-}), \\ h_{2}^{-}c^{+*} =&iJa^{+*}-ig_{N}c_{s}^{*}(B_1^{+*}+B_1^{-})\\ &-ig_{M}c_{s}^{*}(B_2^{+*}+B_2^{-}), \\ h_{3}^{+}B_{1,2}^{-} =&ig_{N,M}(c_{s}^{*}c^{-}+c_{s}c^{+*}), \\ h_{3}^{-}B_{1,2}^{+*} =&-ig_{N,M}(c_{s}^{*}c^{-}+c_{s}c^{+*}).\end{aligned}
		\label{A2}
	\end{eqnarray} 
	Solving Eq.~(\ref{A2}) leads to
	\begin{eqnarray}
		\begin{aligned}
			a^{-} &=\frac{[(h_{2}^{+}\Pi-2i\lambda)(\Pi\Gamma^{-}+2ih_{1}^{-}\lambda)-4h_{1}^{-}\lambda^{2}]\epsilon_{p2}}{(\Pi\Gamma^{+}-2ih_{1}^{+}\lambda)(\Pi\Gamma^{-}+2ih_{1}^{-}\lambda)-4h_{1}^{+}h_{1}^{-}\lambda^{2}},\end{aligned}
		\label{9}
	\end{eqnarray} 
	where $a^{-}$ is the coefficients of the first-order upper sidebands in this case, and we can derive the transmission rate of the probe field, 
	\begin{eqnarray}
		\begin{aligned}
			T_a=|t_{p}|^{2}=|1-\frac{\kappa_{ex2}a^{-}}{\epsilon_{p2}}|^{2}.
			\label{A5}
		\end{aligned}
	\end{eqnarray} 
	
	\section{OMIT and group delay with the broken critical coupling}	
	\begin{figure}[t] 	\centering
		\includegraphics[width=1\linewidth]{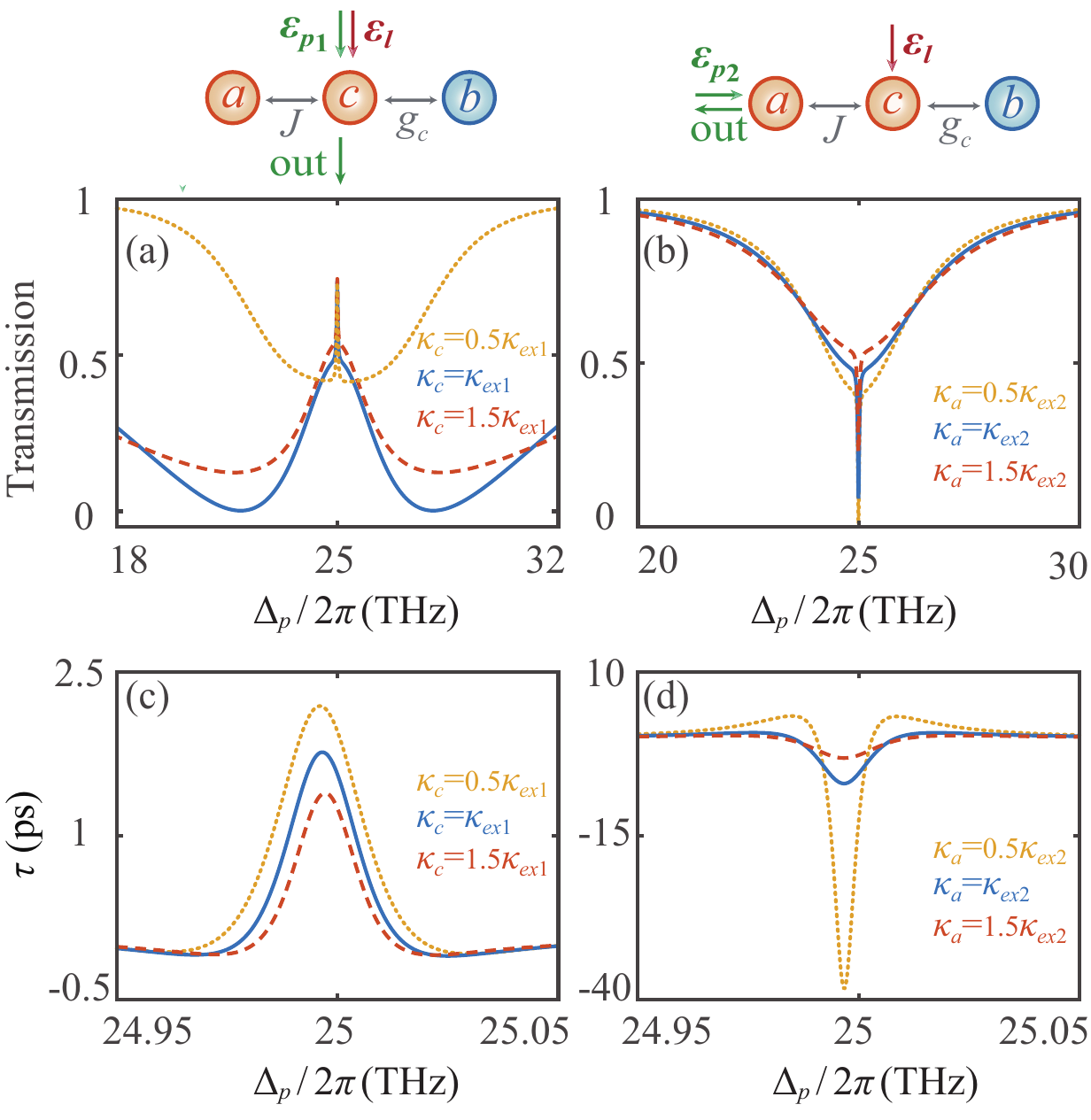}
		\caption{Transmission rate as a function of the optical detuning $\Delta_p$ and intrinsic optical losses $\kappa_{a,c}$ for the input port of the probe field and output port with the plasmonic nanocavity (a), for the input port of the probe field and output port with the microdisk cavity (b). Group delay $\tau$ of the probe field as a function of the optical detuning $\Delta_p$ and intrinsic optical losses $\kappa_{a,c}$ for the input port of the probe field and output port with the plasmonic nanocavity (c), for the input port of the probe field and output port with the microdisk cavity (d). The parameters used here are $N$$\,=\,$$50$ at (a-d) and $P_l$$\,=\,$$0.5\,\mathrm{mW}$ at (c)(d).}
		\label{fig5}
	\end{figure}
	In the main text, we discuss OMIT and group delay under critical coupling conditions. Hence, in this section, we mainly focus on the impact of breaking critical coupling on OMIT and group delay. Figure.~\ref{fig5} predominantly illustrates the transmission rate and group delay as functions of detuned $\Delta_p$ under varying intrinsic optical losses $\kappa_{a,c}$ at different ports of the probe light. Firstly, for the input port of the probe field and output port with the plasmonic nanocavity, under different intrinsic dissipations in the cavity, OMIT spectral lines exhibit similarities and maintain comparable cascaded transparent windows [see Fig.~\ref{fig5} (a)]. Just for the intrinsic optical losses $\kappa_c=0.5\kappa_{ex1}$, the full width at half maximum of the peak is smaller, and from the graph, it seems that there is no cascaded transparent window (see the yellow dotted line). For group delay, when the intrinsic optical losses are smaller, a greater number of photons are present within the cavity, further enhancing the interference effect and leading to a stronger slow light effect [see Fig.~\ref{fig5} (c)].
	
	Likewise, for the input port of the probe field and output port with the microdisk cavity, similar OMIA spectral lines are maintained under varying intrinsic optical losses $\kappa_a$. For different $\kappa_a$, the full width at half maximum of the spectral line and the value of the absorption valley are different [see Fig.~\ref{fig5} (b)], which is due to the influence of intrinsic optical losses on the interference effect. The same applies to the fast light effect [see Fig.~\ref{fig5} (d)].

	\section{Stability analysis}

\begin{figure}[t] 	
		\centering
		\includegraphics[width=0.8\linewidth]{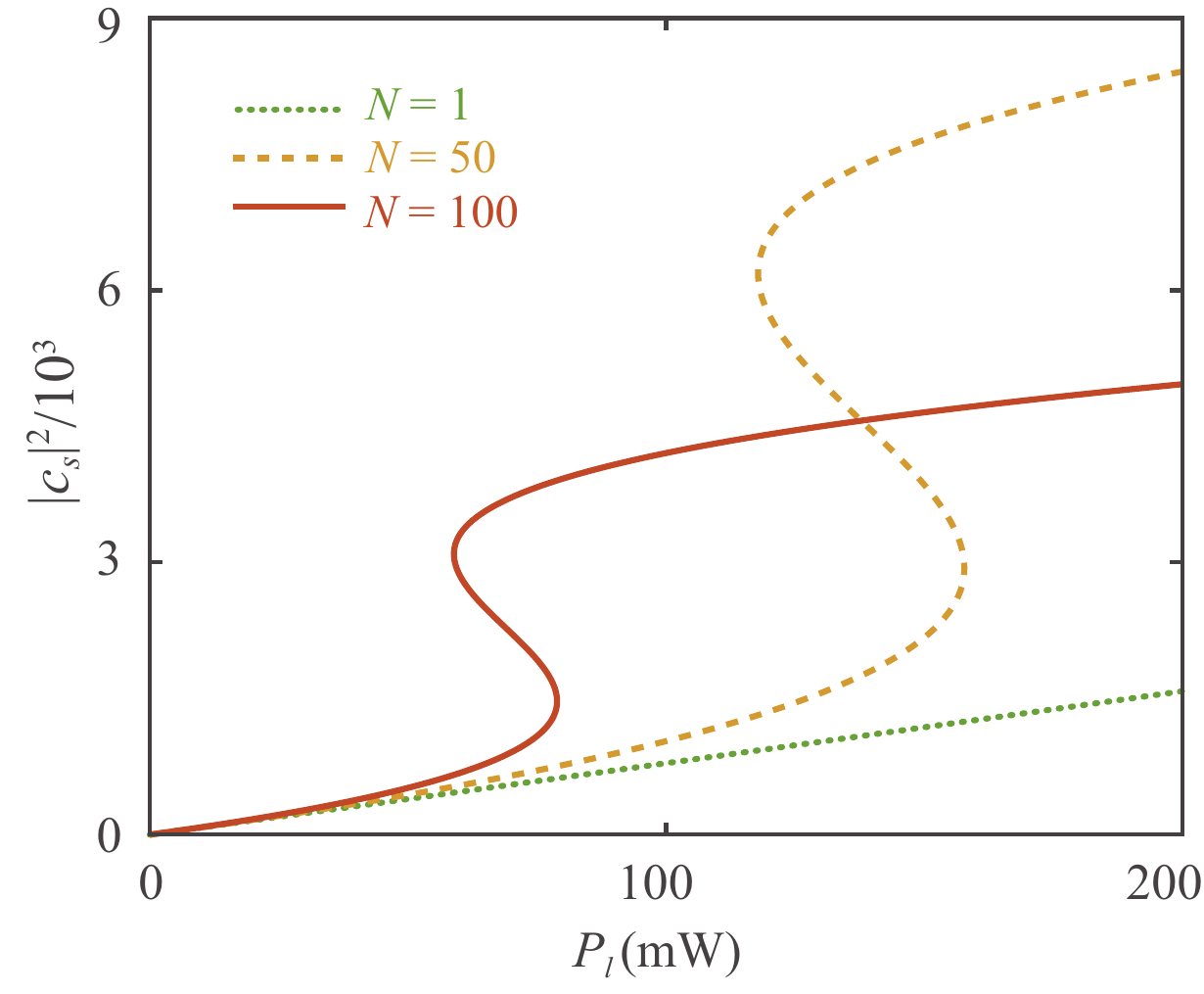}
		\caption{Stability analysis. Mean photon number $|c_s|^2$ as a function of the pump power $P_l$ at different $N$.}
		\label{fig6}
	\end{figure}

	\vspace{-0.5em}
	The steady-state value of this system is
	\begin{eqnarray}
		\begin{aligned}
			a_{s} &=\frac{-iJc_{s}}{i\Delta_{a}+\kappa_{a}},   ~~~c_{s} =\frac{-iJa_{s}+\epsilon_{l}}{i\Delta_s+\kappa_{c}},\\ B_{1s,2s} &=\frac{ig_{N,M}|c_{s}|^{2}}{i\omega_{b}+\gamma}, \end{aligned}
		\label{B1}
	\end{eqnarray}
	where $\Delta_{s}$$\,=\,$$\Delta_{c}-g_{N}(B_{1,s}^{*}+B_{1,s})-g_{M}(B_{2,s}^{*}+B_{2,s})$ including the optical frequency shift induced by molecular vibration. We now study the steady-state behavior of the mean photon number $|c_s|^2$. Moreover, it should be noted that for the different paths that have been considered within the main text, they do not affect the steady-state behavior. By solving the steady-state solution equations above, we can obtain the steady-state photon number $|c_s|^2$ satisfied
	\begin{eqnarray}
		\begin{aligned}
			C_3x^3+C_2x^2+C_1x^1+C_0=0,
			\label{B2}
		\end{aligned}
	\end{eqnarray} 
	where we difine $x$$\,=\,$$|c_s|^2$, and the coefficients are
	\begin{eqnarray}
		\begin{aligned}
			C_{0} =&-\epsilon_{l}^{2}(\Delta_{a}^{2}+\kappa_{a}^{2}),\\C_{3} =&(\nu+\mu)^{2}(\Delta_{a}^{2}+\kappa_{a}^{2}),\\ C_{2} =&2[\Delta_{a}J^{2}(\mu+\nu)-\Delta_{c}(\mu-\nu)(\Delta_{a}^{2}+\kappa_{a}^{2})],\\ C_{1} =&(\Delta_{c}^{2}+\kappa_{c}^{2})(\Delta_{a}^{2}+\kappa_{a}^{2})+2(\kappa_{a}\kappa_{c}-\Delta_{a}\Delta_{c})J^{2}+J^{4},
			\label{B3}
		\end{aligned}
	\end{eqnarray} 
	where $\mu =2g_{N}^{2}\omega_b/(\omega_b^{2}+\gamma^{2})$ and $\nu=g_{M}^{2}\omega_b/(\omega_b^{2}+\gamma^{2})$. Figure.~\ref{fig6} shows the mean photon number $|c_s|^2$ varies the pump power $P_l$ at different $N$ by sovling Eq.~(\ref{B2}) numerically. For $N$$\,=\,$$100$, it can be known that for case $P_l<58\mathrm{mW}$, only one solution exists and the system has no bistability. For case $P_l$ is larger than $58.9\mathrm{mW}$ while less than $78.9\mathrm{mW}$, there are three solutions. So the system gives rise to bistability in this case. For case $P_l>78.9\mathrm{mW}$ at $N$$\,=\,$$100$, the system also has stability. However, when $N$$\,\neq\,$$100$, the system may be in bistability zone for $P_l>78.9\mathrm{mW}$. Thus, to obtain an OMIT, the stability region should be chosen, and we hold $P_l<58mW$ throughout the work.
	
	%
	
\end{document}